\documentclass[conference]{IEEEtran}
\usepackage{amsmath,amssymb,amsfonts}
\usepackage{algorithmic}
\usepackage{graphicx}
\usepackage{textcomp}
\usepackage{xcolor}
\usepackage{balance}
\usepackage[hidelinks]{hyperref}
\usepackage{multirow}
\usepackage{multicol}
\usepackage{listings}
\usepackage{comment}
\usepackage[numbers,sort]{natbib}

\colorlet{punct}{red!60!black}
\definecolor{background}{HTML}{EEEEEE}
\definecolor{delim}{RGB}{20,105,176}
\colorlet{numb}{magenta!60!black}

\lstdefinelanguage{json}{
    basicstyle=\normalfont\ttfamily,
    numbers=left,
    numberstyle=\scriptsize,
    stepnumber=1,
    numbersep=8pt,
    showstringspaces=false,
    breaklines=true,
    frame=lines,
    backgroundcolor=\color{background},
    literate=
     *{0}{{{\color{numb}0}}}{1}
      {1}{{{\color{numb}1}}}{1}
      {2}{{{\color{numb}2}}}{1}
      {3}{{{\color{numb}3}}}{1}
      {4}{{{\color{numb}4}}}{1}
      {5}{{{\color{numb}5}}}{1}
      {6}{{{\color{numb}6}}}{1}
      {7}{{{\color{numb}7}}}{1}
      {8}{{{\color{numb}8}}}{1}
      {9}{{{\color{numb}9}}}{1}
      {:}{{{\color{punct}{:}}}}{1}
      {,}{{{\color{punct}{,}}}}{1}
      {\{}{{{\color{delim}{\{}}}}{1}
      {\}}{{{\color{delim}{\}}}}}{1}
      {[}{{{\color{delim}{[}}}}{1}
      {]}{{{\color{delim}{]}}}}{1},
}

\newcommand\code[1]{{\small \texttt{#1}}}

\def\BibTeX{{\rm B\kern-.05em{\sc i\kern-.025em b}\kern-.08em
    T\kern-.1667em\lower.7ex\hbox{E}\kern-.125emX}}
\usepackage[detect-all,group-separator={,}]{siunitx}

\newcommand{\fabric}[0]{Octopus}
\newcommand{\fabricAbbr}[0]{Octopus}
\newcommand{\service}[0]{Octopus Web Service}
\newcommand{\serviceAbbr}[0]{OWS}

\begin{document}
\title{\fabric: Experiences with a Hybrid Event-Driven Architecture for Distributed Scientific Computing}

\author{
    \IEEEauthorblockN{
        Haochen Pan\IEEEauthorrefmark{1}, 
        Ryan Chard\IEEEauthorrefmark{2}, 
        Sicheng Zhou\IEEEauthorrefmark{3}, 
        Alok Kamatar\IEEEauthorrefmark{1}, 
        Rafael Vescovi\IEEEauthorrefmark{2}, \\ 
        Valérie Hayot-Sasson\IEEEauthorrefmark{1}, 
        André Bauer\IEEEauthorrefmark{4}, 
        Maxime Gonthier\IEEEauthorrefmark{1}, 
        Kyle Chard\IEEEauthorrefmark{1}\IEEEauthorrefmark{2}, 
        Ian Foster\IEEEauthorrefmark{2}\IEEEauthorrefmark{1}
    }
    \IEEEauthorblockA{
        \IEEEauthorrefmark{1}Department of Computer Science, University of Chicago, Chicago, IL, USA\\
    }
    \IEEEauthorblockA{
        \IEEEauthorrefmark{2}Data Science and Learning Division, Argonne National Laboratory, Lemont, IL, USA\\
    }
    \IEEEauthorblockA{
        \IEEEauthorrefmark{3}Department of Computer Science, Southern University of Science and Technology, Guangdong, China\\
    }
    \IEEEauthorblockA{
        \IEEEauthorrefmark{4}Department of Computer Science, Illinois Institute of Technology, Chicago, IL, USA\\
    }
}

\maketitle

\begin{abstract}
Scientific research increasingly relies on distributed computational resources, storage systems, networks, and instruments, ranging from HPC and cloud systems to edge devices. 
Event-driven architecture (EDA) benefits applications targeting distributed research infrastructures by enabling the organization, communication, processing, reliability, and security of events generated from many sources.
To support the development of scientific EDA, we introduce Octopus, a hybrid, cloud-to-edge event fabric designed to link many local event producers and consumers with cloud-hosted brokers, and to provide a fabric for developing resilient applications.
Octopus can be scaled to meet demand, permits the deployment of highly available Triggers for automatic event processing, and enforces fine-grained access control.
We identify requirements in self-driving laboratories, scientific data automation, online task scheduling, epidemic modeling, and dynamic workflow management use cases, and present results demonstrating Octopus’ ability to meet those requirements. 
Octopus supports producing and consuming events at a rate of over 4.2 M and 9.6 M events per second, respectively, from distributed clients.

\end{abstract}


\section{Introduction} \label{sec:intro}
Modern science increasingly requires the coordinated use of advanced computing, networking, instruments, and experimental facilities---collectively, \textit{distributed research infrastructure} (DRI)~\cite{IRI}---extending from HPC systems to high-data-rate instruments and less well-connected edge systems, and often also encompassing cloud-hosted services.
Such DRI can enable a range of important applications, from AI-enhanced instruments~\cite{babu2023deep,pithan2023closing} and self-driving labs \cite{sparkes2010towards,hase2019next,stach2021autonomous,abolhasani2023rise} to epidemic response~\cite{collier2023developing} and scientific data automation~\cite{vescovi2022linking, chard23automation}.

The development of such applications is complicated by the need to address traditional distributed computing challenges, such as handling failures, while also meeting stringent performance requirements and accommodating the unique properties of scientific infrastructure.
Methods used historically to develop scientific applications within single machines or institutions, like message passing and workflow tools, fare badly on dynamic, heterogeneous DRI, in which anomalies are routine rather than unexpected.
At the same time, the event-driven architectures (EDA) used to construct reliable and scalable distributed applications~\cite{michelson2006event,hinze2009event,dayarathna2018recent} struggle with the often large data volumes of scientific applications, and with specialized properties of DRI, from unique event sources and event handling actions to large number of authorization domains and the overhead to provide robust and scalable multi-user capabilities on which scientists can depend.

A promising solution to these challenges is to adapt EDAs to address the specialized requirements of DRI and scientific applications.
Here we report on the investigations that we have undertaken with a view to achieving that goal.
We ask, in particular:
How well do scientific applications map to EDA constructs?
What are the performance characteristics of different implementation approaches?
How can we enable diverse actions to be automatically launched as a result of events?
How feasible is it for EDA to integrate with existing DRI and scientific services to serve the research community?
How can we build a multi-tenant EDA that serves the needs and meets the scale of broad science use cases?

\begin{figure}
\centering
{\includegraphics[page=1,trim={0.5cm, 0cm, 0.28cm, 0.5cm}, clip, width=\columnwidth]{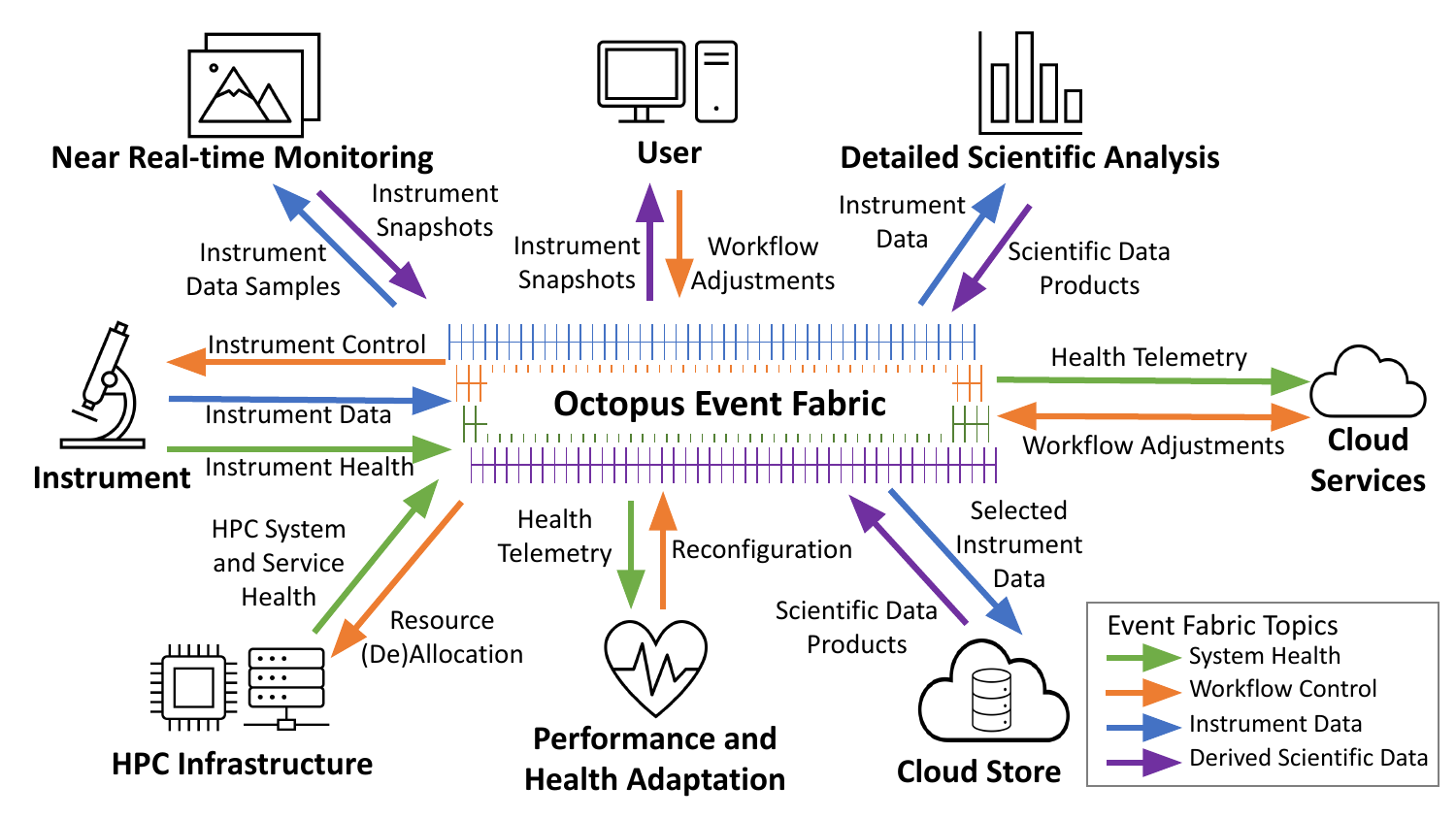}}
\caption{The \fabric{} event fabric spans locations and provides an available and resilient basis for developing applications. Various event topics allow filtered views by consumers.
}
\label{fig:streams}
\vspace{-1.5em}
\end{figure}

As part of our investigation of these questions, we have developed
\fabric{}, a prototype global event fabric and software ecosystem that enables scientific users and applications to communicate and process events within DRI.
We design
\fabric{} around a globally-accessible, cloud-hosted, multi-user service with fine-grained access control, enabling the secure publication and consumption of events from any internet-accessible location: see \autoref{fig:streams}. To address application requirements, we implement functionality in \fabric{} to process events and to \textit{trigger} arbitrary actions (e.g., via web services) across the distributed ecosystem.
\fabric{} users can define rules that specify the actions to take when particular events are received.
For example, a first rule might state that data acquisition at an instrument should trigger a workflow to transfer the data to an HPC system;
a second that completion of the transfer should trigger analysis on the HPC; and a third that conclusion of the analysis should trigger an email to a researcher with results.

We are concerned with both the performance and practical usability of \fabric{}.
To address the first question, we present microbenchmarks that quantify its scalability and performance.
To investigate the second, we examine its use in five applications: a self-driving laboratory, distributed data monitoring and automation, FaaS scheduling, epidemic modeling, and workflow monitoring. In each case, we show how \fabric{}, and more generally, EDA, can enable new functionality not previously possible.

The contributions of our work are:

\begin{enumerate}

  \item Investigation of using scientific EDA to design robust distributed applications, a survey of use cases to determine requirements, and discussion of the benefits and limitations of applying EDA to five science applications.

  \item Design and implementation of a hybrid, cloud-hosted, multi-user event fabric, \fabric{}, along with its open-sourced software ecosystem. The Octopus SDK and a walkthrough notebook are available at \href{https://github.com/globus-labs/diaspora-event-sdk}{github.com/globus-labs/diaspora-event-sdk}.

  \item Evaluation of \fabric{}' performance, scalability, and suitability for scientific use cases. The evaluation methodology and results are detailed at \href{https://doi.org/10.5281/zenodo.10975534}{doi.org/10.5281/zenodo.10975534}
  
\end{enumerate}

The rest of this paper is as follows.
We introduce EDA in \autoref{sec:EDA} and review in \autoref{sec:motivation} scientific application requirements that motivate this work.
We then introduce our key contribution in \autoref{sec:implementation}, the cloud-hosted \fabric{} event fabric, and describe our implementation of this model.
We evaluate \fabric{} in \autoref{sec:evaluation}. In \autoref{sec:application}, we present five applications that leverage \fabric{} and show how our design meets important requirements. In \autoref{sec:discussion}, we discuss our experience building scientific EDA, and discuss the limitations and cost of our approach. Finally, we discuss related work in \autoref{sec:related-work} and summarize in \autoref{sec:summary}.

\section{Event-Driven Architecture}\label{sec:EDA}
Event-driven architecture (EDA) is a software architecture paradigm that revolves around the production, detection, consumption, and reaction to events~\cite{michelson2006event,hinze2009event,dayarathna2018recent}.
EDA is particularly useful in designing decentralized applications that need to be responsive, asynchronous, and reliable, despite dynamic behaviors and unreliable underlying infrastructure.

An \textit{event} represents a state change or significant occurrence within the system. In conventional EDA, events are typically lightweight, containing just the information needed for event handlers to respond appropriately. As we describe below, events in scientific applications may be much larger.

\textit{Event producers} generate and publish events, often without knowing what will eventually consume or process them; \textit{event consumers}
listen for and react to events, for example by updating a database, triggering a process, or generating more events.
Events are transported from producers to consumers via \textit{event channels} implemented in various ways, including message queues, brokers, or event streams. Event channels may be organized into \textit{topics} that buffer related events.

\textit{Triggers} enable \textit{actions} to be performed automatically when specific events, or groups of events, are published.
Using a trigger-action programming model~\cite{huang15tap,ur16trigger}, clients can specify via
rules the actions, such as calls to remote web services, to be run when events meet certain criteria.

EDA has proven popular due to its ability to achieve the following properties.
\textit{Scalability}: EDA can handle high volumes of events and scale out to accommodate growth, as components can be distributed across different systems or networks.
\textit{Flexibility and adaptability}: New event producers and consumers can be added or removed without disrupting the system, making it easy to adapt to changing requirements.
\textit{Asynchronous communication}: Components in an EDA do not directly call each other but communicate asynchronously through events, reducing system coupling and increasing resilience.
\textit{Real-time processing}: EDA is well-suited for applications that require rapid responses, as events can be processed and acted upon as they occur.

EDA challenges include managing event consistency, ensuring reliable event delivery, and handling the complexity of debugging in a distributed system.

\section{Use Cases \& Requirements} \label{sec:motivation}
To leverage EDA concepts in scientific computing, we must first understand the requirements of distributed scientific applications. To this end, we investigate five representative use cases: Self-driving laboratories, scientific data automation, online task scheduling, epidemic modeling and response, and workflow monitoring and steering. In the following, we describe each of these use cases, which have event characteristics as described in~\autoref{table:use-cases}.
Then, based on these use cases, we identify key requirements that must be met by \fabric{}.

\vspace{-0.1in}
\begin{table}[h]
  \caption{Characteristics of events for \fabric{} use cases. $R$ stands for the number of managed resources.}
  \label{extractor-tab}
  \vspace{-0.2in}
  \begin{center}
    \resizebox{\linewidth}{!}{
      \begin{tabular}{llrcccc}
        \hline
        \textbf{Use Case} &
        \textbf{\begin{tabular}[c]{@{}c@{}}Events/\\Hour\end{tabular}} &
        \textbf{\begin{tabular}[c]{@{}c@{}}Mean\\Event\\Size\end{tabular}} &
        \textbf{\begin{tabular}[c]{@{}c@{}}Number\\Topics\end{tabular}} &
        \textbf{\begin{tabular}[c]{@{}c@{}}Number\\Producers\end{tabular}} &
        \textbf{\begin{tabular}[c]{@{}c@{}}Number\\Consumers\end{tabular}} \\
        \hline
        SDL        & $O(10^2)  \times R$      & 0.5~KB & 1   & $R$ & 1       \\
        Data Auto. & $O(10^3)  \times R$    & 4~KB   & 1   & $R$ & Trigger \\
        Scheduling & $O(10^4)  \times R$   & 1~KB   & $R$ & $R$ & 1       \\
        Epidemic   & $O(10)  \times R$ & 1~KB   & $R$ & $R$ & Trigger \\
        Workflow   & $O(10^3)  \times R$    & 1~KB   & $R$ & $R$ & $R$     \\
        \hline
      \end{tabular}
    }
  \end{center}
  \vspace{-2em}
  \label{table:use-cases}
\end{table}

\subsection{Use Cases}

\textbf{Self-driving laboratories} (SDLs)~\cite{seifrid22sdl} aim to transform scientific research by integrating AI, robotics, and HPC resources to continuously and autonomously design, conduct, and analyze experiments with minimal human intervention.
Using an EDA to coordinate experiments across globally distributed and diverse instruments and computational resources that span administrative domains requires accessible and reliable services to communicate events.
Even small SDLs can generate 100s of events per hour that record each workflow step and robot action. Each event includes the name of the instrument, timestamp, experiment identifier, and action description, and may contain associated metadata or results generated by the process. Events are used to prompt new actions in the laboratory; monitor and visualize processes; detect error conditions; and track provenance of an experiment.

\textbf{Scientific data automation} is increasingly necessary to keep pace with the large volumes of distributed data produced in science. Automation relies on capturing, aggregating, and acting on distributed events (e.g., from file systems, instruments, or machine learning (ML) models).
For instance, high-performance filesystems can perform billions of events (e.g., file creation, modification, deletion) per day~\cite{gray2005scientific}. Centralized workflow methods~\cite{chard23automation}  have been used for similar purposes but require static definitions of processing pipelines, while rule-based storage systems~\cite{rajasekar2010irods} focus only on file management operations. Applying an EDA would enable data management policies to be expressed as simple trigger-action rules, for example, performing analysis workflows, ML training or inference, or backup procedures upon data modification or arrival. However, this would require event producers to be applicable across a diverse range of research infrastructure.

\textbf{Online task scheduling} is used to optimally distribute workloads across computing resources based on various metrics and goals (e.g., cost, resource usage, power consumption, and available resources).
Intelligent scheduling not only benefits users, it also ensures that computational resources are used efficiently, leading to improved system performance and reliability.
Especially for federated serverless computing environments~\cite{chard2020funcx} that link distributed computing resources (from edge devices to supercomputers), it is crucial to have near-real-time and fine-grained information about task performance and resource usage. This data can then be used for building models and for online decision making.
This necessitates near real-time monitoring to capture tens of thousands of events each hour for every participating resource and thus requires event infrastructure to scale accordingly.

\textbf{Epidemic modeling and response} gained much importance as a decision-making tool during the COVID-19 pandemic. Researchers preparing for future pandemics are building automated systems to collect and analyze data rapidly to inform decision makers~\cite{collier2023developing}. Such approaches rely on streams of data collected from both instrumentation (e.g., public health information, hospitals, mobility data, social media~\cite{o2017digital}) and trained models. The processing of this data necessitates cleaning and validation, executing analyses for transformation into a common schema, visualization and distribution, triggering of new analyses (such as computing R values), and notifying decision makers on observed or predicted trends.  Similar requirements exist in smart cities~\cite{strohbach2015towards, catlett20aot} and modeling and response to natural disasters~\cite{altintas23wifire}.

\textbf{Dynamic workflow management} plays a pivotal role in orchestrating the execution of complex workflows across (distributed) resources. Resource deficits, whether caused by increased demand or hardware failure, can compromise overall progress or lead to inefficient execution. In the worst-case scenario, critical tasks may remain unperformed, leading to
failure of the entire workflow. As a consequence, it is necessary to monitor the performance of the workflow and the utilized resources to detect issues and implement adaptive healing actions before they escalate into failures~\cite{li22faulttolerant}. These actions may involve dynamically reconfiguring workload partitioning (e.g., assign less work to stragglers) and/or routing parts of the workflow to different resources.

\subsection{Requirements}
Although our use cases have different objectives and characteristics (see \autoref{extractor-tab}), they share common requirements, including the following.

\textbf{Scalability}: Our use cases can exhibit 
peak data rates exceeding \num{10000} events per minute. \fabric{} must be able to efficiently serve these requests, and to must support many users and applications concurrently.

\textbf{Fault tolerance}: In order for \fabric{} to function as essential infrastructure for scientific purposes, it needs to exhibit reliability and resilience against failures. This means ensuring the delivery of events even in the the case of messaging broker or client interruptions. It should also provide clients with a straightforward interface for managing failures.

\textbf{Diversity of event schemata}: Each  use case includes multiple message types with differing structures and payloads. While there are common metadata across events, such as timestamps, the majority of the message types are unique and require that \fabric{} accommodates a flexible event schema.

\textbf{Usability}: Event publishers and consumers may run on Cloud, Fog, Edge, and HPC resources, utilizing different programming languages. The applications themselves could range from simple programs to components within complex frameworks, frontend user interfaces, storage connectors, and other forms, whether centralized or distributed.
It is crucial that \fabric{} supports connections for all these variations.

\textbf{Triggers}: A common requirement in several use cases is to perform a particular action (e.g., invoke a web service to transfer data) whenever an event is observed that satisfies some specified criteria.
While a user could subscribe to the relevant topic and implement this behavior themselves locally, it can be both simpler and more efficient for \fabric{} to allow the user to register an action that is then applied automatically to any event that meet a user-defined condition.

\textbf{Performance}
should be predictable and scalable, exhibiting a sufficiently low latency when transmitting and receiving events. In addition, \fabric{} must only minimally increase communication costs as many of the use cases rely on the efficient transfer of information between publishers and consumers to steer applications in near-real-time.

\textbf{Fine-grained access control}: Communication with \fabric{} must be authenticated, authorized, and auditable. Each user or a group of users must be allowed to access only their topics, such that users cannot consume or produce events to unauthorized topics. Users require the ability to self-manage access control on their topics.

\section{Design and Implementation} \label{sec:implementation}
\begin{figure}
  \centering
  \includegraphics[page=2, width=\linewidth, trim={4cm 1.2cm 4.7cm 1.2cm}, clip]{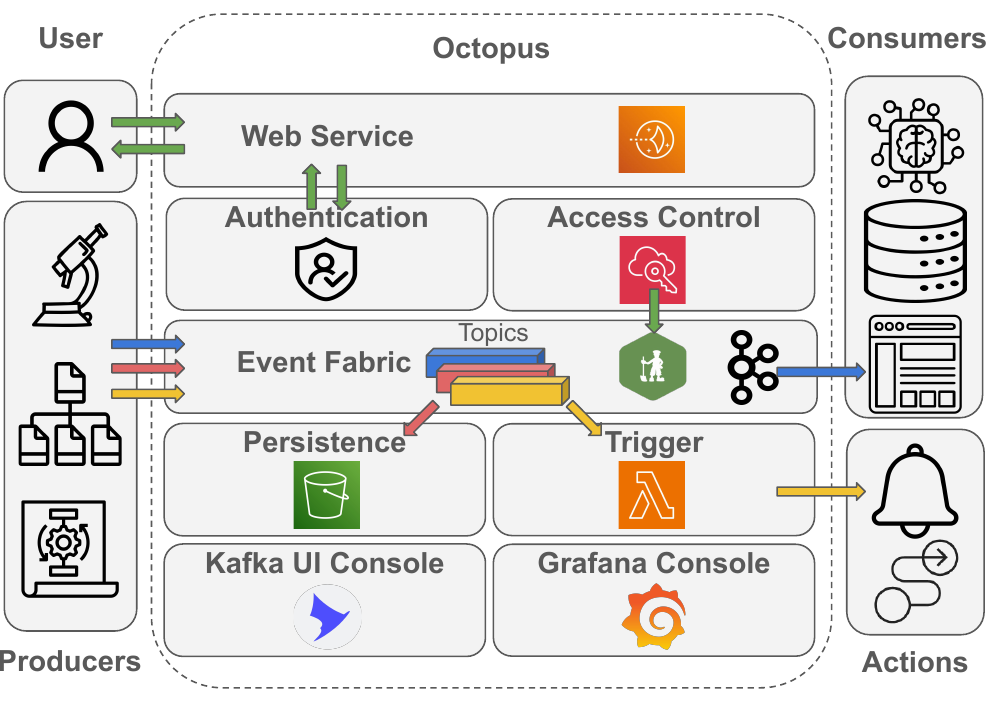}
  \caption{\fabric{} architecture. Users interact with the web service (green arrows) to acquire credentials, where their identity is mapped to an IAM entity and shared with the MSK ZooKeeper. Producers (left) and consumers (right) communicate with the event fabric to publish and receive events, respectively (blue arrows). A trigger can be configured to act on events and invoke remote actions (yellow arrows). Events can also be persisted to reliable cloud storage when enabled (red arrows). Monitoring consoles are available for admins to monitor the system's live status and historical statistics.}
  \label{fig:architecture}
  \vspace{-1.5em}
\end{figure}

\fabric{} is a hybrid edge-cloud EDA platform that combines a cloud-hosted web service and messaging fabric with remote producer and consumer clients, as shown in \autoref{fig:architecture}.
Here we describe these components and explain how users can use \fabric{} to implement scientific EDAs.

\subsection{Event Fabric}

We build the \fabric{} event fabric on Apache Kafka,
an event streaming platform with high throughput, reliability, and scalability~\cite{kreps2011kafka}.
A Kafka deployment runs on a cluster of one or more servers called \textit{brokers}, which expose a publish/consume interface and handle the receipt of messages from producers and their dispatch to consumers. Messages are grouped first into topics and then partitions, with messages within a partition strictly ordered.
Large Kafka clusters, with more than \num{4000} brokers, have been used to process over 7 trillion messages daily~\cite{linkedin2019how}. Empowered by Kafka, \fabric{} provides a substrate for both simple and complex workflows through general eventing interfaces, making it ideally suited for complex, real-time, asynchronous workflows.

Rather than host our own Kafka cluster, we use AWS Managed Streaming for Kafka (MSK)---a highly available, scalable, and robust Kafka deployment that benefits from integration with other AWS services, such as IAM identity management, S3 for persistent storage, and Amazon CloudWatch for monitoring. MSK allows the \fabric{} event fabric to be scaled dynamically and geo-replicated for reliability across AWS service regions, while enforcing fine-grained access control that can restrict access on a per-topic level for authenticated users. One potential downside, discussed below, is the costs associated with this use of AWS services.

\subsection{\fabric{} Web Service}
The \service{} (\serviceAbbr{}) allows users to provision, configure, and manage topics; acquire and manage credentials to access MSK topics; and deploy and manage triggers.
\serviceAbbr{} is implemented as a RESTful service
deployed on AWS Lightsail instances.
\serviceAbbr{} implements a comprehensive security model, using OAuth 2.0, via which users must first authenticate via Globus Auth~\cite{tuecke2016globus}, and include access tokens in subsequent API requests.
\serviceAbbr{} exposes REST routes
to provision, configure, share, or release topics:

\begin{itemize}
  \item
        \code{PUT <service>/topic/<topic>}:
        Registers a unique topic name with the MSK cluster and sets \texttt{READ}, \texttt{WRITE}, and \texttt{DESCRIBE} access to the topic for the user identity.

  \item
        \code{GET <service>/topics}:
        Returns a list of all topics the user is authorized to describe.

  \item
        \code{GET <service>/topic/<topic>}:
        Returns a specific topic's configuration.

  \item
        \code{POST <service>/topic/<topic>}:
        Set topic configuration, e.g., replication factor and data retention policy.

  \item
        \code{POST <service>/topic/<topic>/partitions}:
        Set the number of partitions for the topic.

  \item
        \code{POST <service>/topic/<topic>/user}:
        Grant (or revoke) an identity access to the topic.

\end{itemize}

\subsection{Authentication and Authorization}
Authentication and authorization of event producers are important for provenance and accountability and for event consumers to protect sensitive information.
We build our authentication and authorization model on Globus Auth, as it: 1) is a standards-compliant OAuth~2.0 implementation; 2) supports a wide range of identity providers, including many commonly used in research; 3) implements a unique delegation model via which users can delegate permission for services to call other services (e.g., for an external service to register an event in \fabric{}, or for \fabric{} triggers to call external actions, on behalf of a user); and 4) is used by many science services, simplifying their integration as actions in \fabric{}.

As MSK supports only AWS IAM and SCRAM authorization, the MSK cluster uses Zookeeper~\cite{hunt2010zookeeper,zookeeper}, and \serviceAbbr{} uses Globus Auth, it must act as an authorization intermediary between Globus Auth, Amazon IAM authorization, and MSK.
To that end,
\fabricAbbr{} is registered as a Globus Auth \textit{resource server} so that users can authenticate using thousands of supported identity providers.
MSK employs Apache ZooKeeper to maintain and synchronize state (e.g., topics and access control lists) among cluster resources.
Thus, to interact with a topic, a user must be assigned a unique identity and acquire appropriate credentials. 
To achieve this goal, \serviceAbbr{} exposes a route to perform this function:

\begin{itemize}
  \item
        \code{GET <service>/create\_key}:
        Create an IAM identity for the requesting user and return an access key and secret. The IAM identity is registered with the MSK ZooKeeper such that the key and secret can be used to produce and consume events from authorized topics.
\end{itemize}

\subsection{\fabric{} Triggers}
A trigger allows a \fabric{} application to respond to specified events by acting on distributed resources.
Triggers need to be \textit{robust}, meaning that failures can be detected and (where appropriate) actions retried; \textit{scalable}, meaning that many triggers can be run at once; \textit{polyvalent}, meaning that they can perform many different actions; and \textit{empowered}, meaning that they are enabled to act on behalf of authorized users.

We employ the function-as-a-service (FaaS) paradigm to implement a managed trigger model with these capabilities. This approach eliminates the need for users to define, deploy, or manage their own software for responding to events.  Specifically, we deploy, for each trigger, an AWS Lambda function to process events from the associated topic and invoke specified actions. Users can specify the Lambda function and execution environment for that function. We use AWS EventBridge \textit{event patterns}~\cite{eventbridgepatterns} to perform optional event filtering to trigger the Lambda function only on events of interest. EventBridge patterns use a simple JSON-based language for specifying rules
(e.g., see~\autoref{lst:eventbridge})
.
When a user registers a trigger, \serviceAbbr{} configures the optional pattern and also creates the appropriate IAM policy, IAM role, and CloudWatch log group to manage and monitor the Lambda function.

\begin{lstlisting}[language=json, numbers=none, label={lst:eventbridge},
captionpos=b,
basicstyle=\footnotesize\ttfamily,
caption={EventBridge pattern that will invoke the Trigger only when an event with ``event\_type'' is ``created''.}]
[
 {
   "Pattern": "{\"value\":
     {\"event_type\": [\"created\"]}}"
 }
]
\end{lstlisting}

Each Lambda function is given its own MSK \textit{consumer group}, meaning that many instances of the Lambda function can retrieve events without affecting other consumers of the topic.
Lambda functions can scale automatically by evaluating \textit{processing pressure} (the number of pending events in a topic).
Lambda evaluates the processing pressure at 1~min intervals, and scales concurrent invocations of the function dynamically when warranted.
The user can configure (via \serviceAbbr{}) the function to process batches of up to \num{10000} events (or a total of 6~MB) per invocation. \fabric{} Triggers are versatile, secure, robust, and automated, providing a new way for developers to build scientific applications.
Triggers can be managed by:

\begin{itemize}
  \item
        \code{PUT <service>/trigger/}:
        Deploy a trigger using a specified function, target topic, and configuration.

  \item
        \code{GET <service>/triggers/}:
        Describe existing triggers and their configuration.

  \item
        \code{POST <service>/trigger/<trigger\_id>}:
        Update configurations of a function trigger, including batch size, batch window, and filtering criteria.
\end{itemize}

\subsection{\fabric{} SDK}
We provide a Python SDK to interact with the \service{} and simplify integration with existing applications and services.
The SDK is available on PyPI and includes a Globus Auth login manager to perform an authentication flow and cache tokens on the user's behalf. Tokens and MSK secrets are stored in a local SQLite database and automatically refreshed as needed.
Optionally, the SDK can install \texttt{kafka-python} to interact with the event fabric and produce/consume events.
Additionally, other Kafka producer/consumer libraries that support IAM authentication can also use these stored tokens to interact with the \fabric{} cluster.

\subsection{Scalability and Fault-tolerance}
Each \fabric{} component is designed to be independently and dynamically scalable, adapting to real-time demand and scheduled scaling activities. The MSK cluster can be scaled up by resizing cloud instances and by adding additional brokers for horizontal scale up, and fault tolerance can be improved by replicating the cluster across regions.
Topics may be replicated and synchronized by using the Kafka MirrorMaker tool.
Managed consumer functions dynamically scale according to the processing pressure, which is reevaluated each minute to apply the scaling decisions.
\service{}, deployed on Lightsail instances, can be horizontally scaled behind an elastic load balancer.
API operations on the \serviceAbbr{} side are programmed to be idempotent such that the automatic retry of the function would not cause the system to be in inconsistent states. The source of truth about which topics are owned by which identities are stored in ZooKeeper and replicated to IAM, meaning that scaling \serviceAbbr{} does not affect data consistently. Although ZooKeeper provides strong consistency for topic ownership, ownership
updates are infrequent, so there is no bottleneck, and only authenticated users trigger updates through \serviceAbbr{}.

By default, all messages in a topic are stored for seven days. Consumers can consume messages either from the latest or the earliest offset, or after a certain timestamp.
Users can also configure the compaction and retention policy through \serviceAbbr{}.

In order to manage intermittent communication errors robustly, the SDK producer retries a configurable number of times before failing.
Clients can configure the number of acknowledgments required for sending to be considered successful. By default, consumers periodically commit consuming offsets, which provides an at-least-once delivery guarantee. The commit window is adjustable and consumers can manually invoke the commit API.

\section{Evaluation} \label{sec:evaluation}
We investigate the performance and scalability of \fabric{}  using synthetic workloads and micro-benchmarks from the use cases described in \autoref{sec:motivation}.

\subsection{Experimental Setup}
\begin{table}
  \caption{Testbed cluster configurations.}
  \vspace{-2em}
  \begin{center}
    \resizebox{\linewidth}{!}{
      \begin{tabular}{ccccc}
        \hline
        \textbf{Name} & \textbf{\begin{tabular}[c]{@{}c@{}}Number\\ Brokers\end{tabular}} & \textbf{\begin{tabular}[c]{@{}c@{}}Broker\\ Type\end{tabular}} & \textbf{\begin{tabular}[c]{@{}c@{}}vCPUs/\\ Broker\end{tabular}} & \textbf{\begin{tabular}[c]{@{}c@{}}Mem/\\ Broker\end{tabular}} \\
        \hline
        Baseline      & 2                                  & \texttt{kafka.m5.large}            & 2                                  & 8 GB                               \\
        Scale-up      & 2                                  & \texttt{kafka.m5.xlarge}           & 4                                  & 16 GB                              \\
        Scale-out     & 4                                  & \texttt{kafka.m5.large}            & 2                                  & 8 GB                               \\
        \hline
      \end{tabular}
    }
  \end{center}
  \vspace{-2em}
  \label{table:cluster-info}
\end{table}

We use three different cluster configurations for \fabric{}, as described in~\autoref{table:cluster-info}. The first, which we refer to as \textit{baseline}, consists of a two-node MSK cluster in the AWS us-east-1 region using \texttt{kafka.m5.large} instances, each with two virtualized Intel Xeon 8000 CPUs.
The nodes are located in \texttt{us-east-1a} and \texttt{us-east-1b} respectively.
For the second configuration, \textit{scale-up}, uses two nodes of type \texttt{kafka.m5.xlarge}.
For the final configuration, \textit{scale-out}, we use four nodes of type \texttt{kafka.m5.large}.

\begin{table*}[!th]
\centering
\caption{Baseline performance and scalability benchmarking results. Producer and consumer throughputs are events/sec; latency is in milliseconds (ms). A `-' signifies no change from the above result.}
\vspace{-0.5em}
\label{table:evaluation}
\begin{tabular}{cccccc|rrrr|rrrr}
\hline
\multicolumn{6}{c|}{\textbf{Configuration}} & \multicolumn{4}{c|}{\textbf{Local Client}} & \multicolumn{4}{c}{\textbf{Remote Client}} \\ \hline
\textbf{\begin{tabular}[c]{@{}c@{}}Exp.\\ Index\end{tabular}} & \textbf{Cluster} & \textbf{\begin{tabular}[c]{@{}c@{}}Rep.\\ Factor\end{tabular}} & \textbf{\begin{tabular}[c]{@{}c@{}}Num. of\\ Partitions\end{tabular}} & \textbf{\begin{tabular}[c]{@{}c@{}}Prod.\\ Acks\end{tabular}} & \textbf{\begin{tabular}[c]{@{}c@{}}Event\\ Size\end{tabular}} & \multicolumn{1}{c}{\textbf{\begin{tabular}[c]{@{}c@{}}Prod.\\ Thru.\end{tabular}}} & \multicolumn{1}{c}{\textbf{\begin{tabular}[c]{@{}c@{}}Med.\\ Lat.\end{tabular}}} & \multicolumn{1}{c}{\textbf{\begin{tabular}[c]{@{}c@{}}99\%\\ Lat.\end{tabular}}} & \multicolumn{1}{c|}{\textbf{\begin{tabular}[c]{@{}c@{}}Cons.\\ Thru.\end{tabular}}} & \multicolumn{1}{c}{\textbf{\begin{tabular}[c]{@{}c@{}}Prod.\\ Thru.\end{tabular}}} & \multicolumn{1}{c}{\textbf{\begin{tabular}[c]{@{}c@{}}Med.\\ Lat.\end{tabular}}} & \multicolumn{1}{c}{\textbf{\begin{tabular}[c]{@{}c@{}}99\%\\ Lat.\end{tabular}}} & \multicolumn{1}{c}{\textbf{\begin{tabular}[c]{@{}c@{}}Cons.\\ Thru.\end{tabular}}} \\ \hline
1 & Baseline & 2 & 2 & 0 & 32~B & 4,289~K & 54 & 165 & 9,840~K & 4,202~K & 86 & 198 & 9,646~K \\
2 & Baseline & 2 & 2 & 0 & 1~KB & 195~K & 40 & 181 & 356~K & 174~K & 76 & 189 & 367~K \\
3 & Baseline & 2 & 2 & 1 & 1~KB & 161~K & 49 & 179 & - & 143~K & 92 & 209 & - \\
4 & Baseline & 2 & 2 & all & 1~KB & 65~K & 141 & 273 & - & 65~K & 138 & 280 & - \\
5 & Baseline & 2 & 2 & 0 & 4~KB & 43~K & 37 & 146 & 91~K & 39~K & 66 & 174 & 94~K \\
6 & Baseline & 2 & 4 & 0 & 1~KB & 202~K & 32 & 291 & 374~K & 179~K & 73 & 213 & 389~K \\
7 & Scale-up & 2 & 4 & 0 & 1~KB & 238~K & 16 & 352 & 751~K & 184~K & 67 & 279 & 597~K \\
8 & Scale-out & 2 & 4 & 0 & 1~KB & 319~K & 19 & 168 & 785~K & 303~K & 41 & 186 & 813~K \\
9 & Scale-out & 4 & 4 & 0 & 1~KB & 246~K & 27 & 203 & 777~K & 235~K & 47 & 336 & 806~K \\ \hline
\end{tabular}
\vspace{-1.5em}
\end{table*}

Our experiments involve both \textit{local} and \textit{remote} producers and consumers. \textit{Local} producers and consumers are hosted on two EC2 \texttt{c5.24xlarge} instances, each with 96 virtual CPU cores (virtualized Intel Xeon Cascade Lake processor) and 192 GB of memory, located in \texttt{us-east-1a} and \texttt{us-east-1b}, respectively. \textit{Remote} producers and consumers are deployed on two bare-metal Cascade Lake instances on Chameleon Cloud at the Texas Advanced Computing Center (TACC) site. Each instance includes 96 CPU cores (2 Intel Xeon Gold 6240R CPUs) and 192 GB of memory.
The median round-trip time (RTTs) of data between remote instances and MSK instances in this configuration is 46--47 ms, with $<$0.1\% deviation.

\subsection{Data Collection Process}

We implemented a benchmarking operator to orchestrate the creation of topics with specific configurations (e.g., replication factor, number of partitions) and spawn the specified number of producers and consumers on remote resources.
When conducting consumer throughput tests, we first populate the topic with events and then initiate consumers.
This way, all consumers can start from the first offset of the topic and consume events at their own pace.
After an experiment run, the logs from the remote agents are gathered, checked for failures, and aggregated for our analysis.
Each experiment was performed at least three times.
To support the asynchronous tests, we compute the throughput (in event/s) as $T = N / (t_2 - t_1)$; where $N$ is the total number of events and $t_1$ and $t_2$ represent the earliest and latest timestamps, respectively, across all agents, when a second was produced or consumed.
We report the producer's median and 99th percentile latencies as the mean values of each round.
To optimize system throughput and latency, we reduce the producer's \texttt{buffer.memory} to 256~KB and increase the consumer's \texttt{receive.buffer.bytes} to 2~MB.

\begin{figure}
  \centering
  \includegraphics[width=\columnwidth,trim=2.5mm 2.5mm 3mm 2.5mm,clip]{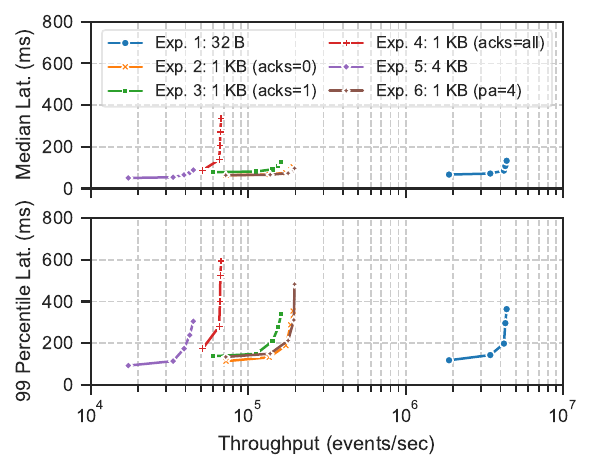}
  \caption{Median (top) and 99th percentile (bottom) latencies vs.\ throughput for configurations 1--6 on baseline cluster with remote producers.
  }
  \label{fig:base-cc}
  \vspace{-1.5em}
\end{figure}

\subsection{Baseline Performance}

We examine producer and consumer performance while varying parameters such as message size, acknowledgments, and number of topic partitions, with results in \autoref{table:evaluation}.

\textbf{Message size:}
In this first set of experiments, we select three message sizes that reflect our use case requirements: 32~B, 1~KB, and 4~KB.
We set the number of acknowledgments to zero (i.e., producers transmit events without waiting for acknowledgment), and both the number of partitions per topic and topic replication factor to two.
The results of these experiments (\#1, \#2, \#5 in \autoref{table:evaluation}) are in \autoref{fig:base-cc}. Each colored line represents an experiment set with 20, 40, 60, 80, and 100 producers. For instance, with 32~B messages (blue line), \fabric{} handles a peak throughput of 4.20 million events per second.
At this throughput, the median latency for local AWS producers is 54~ms, compared to 86~ms for Chameleon Cloud remote producers.
With increasing event sizes, the peak throughput decreases, to 174~K and 39~K event/s for 1~KB and 4~KB events, respectively.

When comparing the throughput of producers and consumers (see \autoref{table:evaluation}), we observe that both local and remote clients exhibit roughly twice the read throughput compared to the write throughput in each configuration.

\textbf{Acknowledgments:}
Experiments \#3 and \#4 change the acknowledgment requirements from 0 to 1 (more precisely, the leading one), and finally, all brokers must acknowledge, while keeping the event size constant at 1~KB.
The peak throughput drops from 195~K (acks=0)
to 161~K (acks=1) down to 65~K (acks=all) event/s for local producers, and from 174~K to 143~K down to 65~K event/s for remote producers.
The median latency increases by 9~ms and 101~ms as well as 16~ms and 62~ms for the local and remote producers.

\textbf{Partitions}:
We increased the number of partitions from 2 to 4 in experiment \#6, while having a fixed size of 1~KB and zero acknowledgments.
The peak throughput increases slightly from 195~K to 202~K event/s for local producers. The median latency decreases slightly, but the 99 percentile latency increases substantially, from 181~ms to 291~ms.

\textbf{Scaling}: We scaled the cluster configuration and changed the event replication factor from 2 to 4, while keeping the event size fixed at 1~KB, and setting acknowledgments to 0 and the number of partitions to 4.
Experiment \#7 shows that the \textit{scale-up} (i.e., doubling the size of the instances) cluster can handle a rate of 238~K and 184~K event/s for remote and local producers, respectively: increases of 17.8\% and 2.8\%, respectively, compared to the \textit{baseline} cluster.
Further, median latency is reduced for both local and remote producers.
For experiment \#8, the \textit{scale-out} cluster (with double the number of instances)
achieves a greater than 50\% increase in peak throughput compared to the \textit{scale-up}
cluster, with 319~K and 303~K event/s for local and remote producers, respectively, and median latency being reduced from 67~ms to 41~ms for remote producers, respectively.

Changing the replication factor from 2 to 4 for \textit{scale-out} (experiment \#9), reduces write throughput from 319~K to 246~K event/s, while read throughput barely changes (785~K to 777~K event/s), for local producers and consumers, respectively.

\textbf{Summary}:
Our experiments show that \fabric{} provides high throughput for small events and that we can increase throughput for larger events at the cost of latency.
We found that the AWS-hosted cluster provided predictable and consistent performance across experiments, and that scaling out is an effective mechanism to accommodate increased demand.
Our experiments validate our selection of technologies and demonstrate \fabric{}' ability to implement EDAs for our use cases and the broader scientific community.

\subsection{\fabric{} Triggers}
We investigate trigger performance in terms of both throughput and scalability.
We observe that trigger throughput increases with additional partitions, larger batch sizes, and reduced data sizes. For instance, if a topic only has 1 partition, consumers of \fabric{} triggers reach 22~K, 7~K, and 2~K event/s throughput for events of size 32~B, 1~KB, and 4~KB, respectively. However, with 8 partitions, the throughput increases, reaching $\sim$147~K, 39~K, and 12~K event/s, respectively, which is roughly six times faster.

\begin{figure}
  \centering
  \includegraphics[width=\columnwidth]{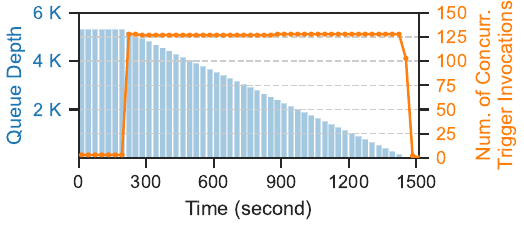}
  \vspace{-1.5em}
  \caption{Trigger scaling using a topic with 128 partitions to receive individual messages and sleep for 30~s. Messages are slowly processed until the processing pressure results in 128 concurrent functions, then scaling down shortly before the workload is complete.}
  \label{fig:trigger-128partitions}
\end{figure}

To investigate trigger scaling, we create a synthetic workload of more than \num{5000} tasks, each sleeping for 30~s. We buffer events evenly across 128 partitions and set the consumer batch size to 1. As shown in \autoref{fig:trigger-128partitions}, the number of trigger consumers is scaled up from 3 to 128 within four minutes and then scaled down shortly before all tasks are complete, showing that Lambda efficiently manages processing pressure.

\begin{figure}
  \centering
  \includegraphics[width=\columnwidth,trim=1mm 1mm 1mm 1mm,clip]{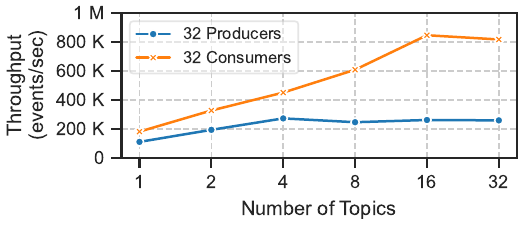}
  \caption{Producer and consumer throughput vs.\ number of topics.
    For producer, increases to 273~K $\times$ 1~KB event/s at four topics and then remains flat. For consumer, increases to 846~K $\times$ 1~KB event/s at 16 topics.}
  \label{fig:multi-tenant}
  \vspace{-1.5em}
\end{figure}

\subsection{Multi-Tenancy}
To evaluate \fabric{} support for multi-tenancy, we measure performance when using multiple topics on 4 brokers in the \textit{scaled-out} cluster,  using 2 clients deployed on AWS instances.
Unlike previous tests, each topic has 1 partition and a replication factor of 2. We increase the number of topics from 1 to 32 in powers of 2. We fix the message size to 1~KB and employ 16 producers, or consumers, per client machine, for a total of 32 producers or consumers in each trial.
Our results, in
\autoref{fig:multi-tenant}, show that producer throughput increases up to 4 topics and then remains flat. The increase from one to four topics is because more brokers can lead producer writes. And since 32 producers already saturate the four brokers, adding more topics does not increase the throughput. Consumer throughput, however, increases until 16 topics.

\section{Applications} \label{sec:application}
The five use cases described earlier have implemented a scientific EDA using \fabric{}. We describe how these use cases have leveraged \fabric{} and summarize lessons learned.

\subsection{Self-driving Laboratories}

Here we tackle the challenge of monitoring the many different events generated by an SDL at Argonne National Laboratory.
The SDL uses \fabric{} to create a global log of distributed actions spanning robotic devices, HPC resources, and data resources. Events include the initiation of experiments, transitions between stages of the workflow, and the collection of results, across various instruments and computational resources. This log facilitates transparent and real-time insights into ongoing experiment workflows as well as enables researchers to trace back through the decision-making and experiment processes for review or optimization. Furthermore, this detailed provenance is invaluable for ensuring reproducibility by recording the lineage of data and will eventually serve as the basis for event-driven automation. The event log is consumed to provide graphical representations of the experiment, allowing administrators to quickly evaluate throughput, understand the stage of various workflows, and inspect event-level information.

\begin{figure*}
\centering
{\includegraphics[page=3, width=0.34\linewidth, trim=0cm 7.9cm 8.8cm 0cm,clip]{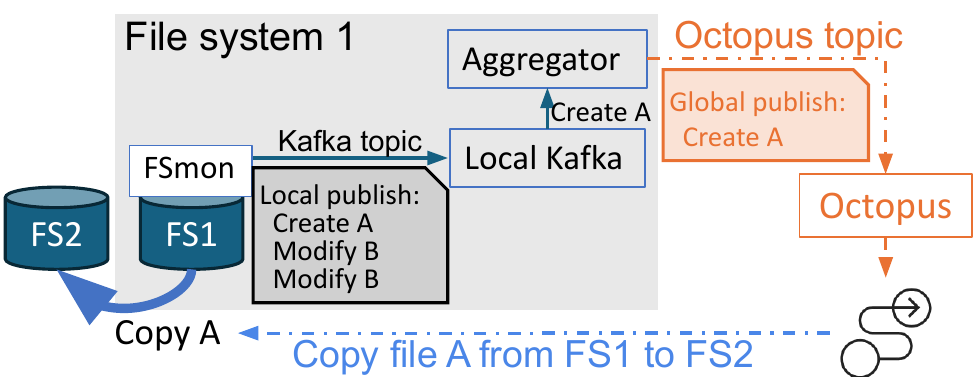}}
\hfill
{\includegraphics[page=5, width=0.30\linewidth, trim=0cm 8.25cm 11cm 0cm,clip]{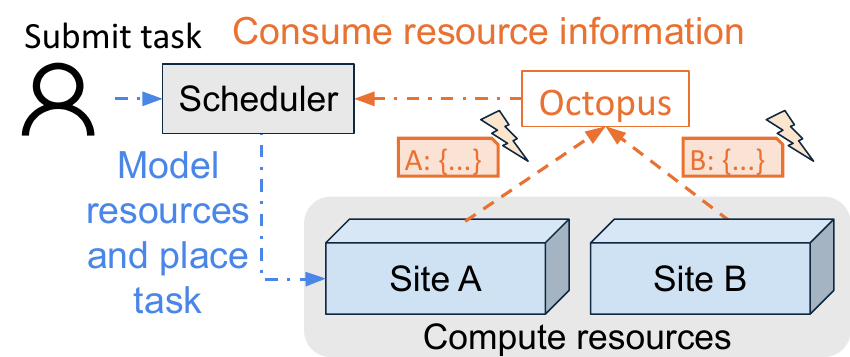}}
\hfill
{\includegraphics[page=4, width=0.34\linewidth, trim=0cm 8.05cm 8.7cm 0cm,clip]{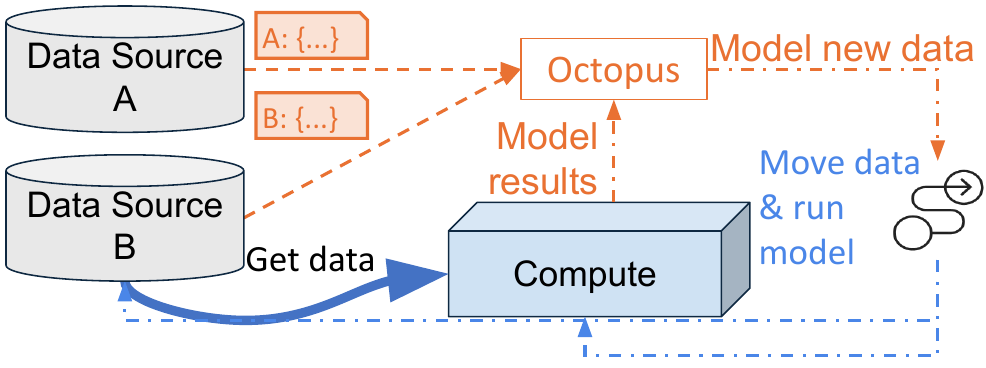}}
\vspace{-0.5em}
\caption{Three scientific EDAs implemented for our use cases: file system synchronization (left), with file creation events from one FS triggering data replication requests (see \autoref{sec:SDA}); online task scheduling (middle), with events carrying energy information to guide scheduling decisions (see \autoref{sec:ots}); and epidemic modeling (right), with events from data sources triggering model inference on new data (see \autoref{sec:epi}).}
\vspace{-1em}
\label{fig:implemented-edas-arch}

\end{figure*}

\subsection{Scientific Data Automation}\label{sec:SDA}

\begin{figure}
  \centering
  \includegraphics[width=\columnwidth,trim=1mm 1mm 1mm 0mm,clip]{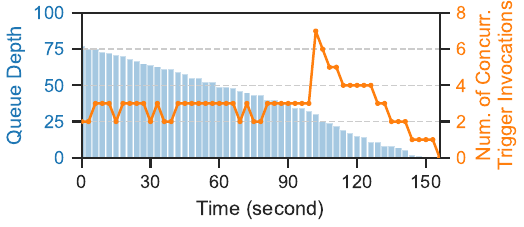}
  \caption{\fabric{} activities for Scientific Data Automation use case.
  Events accumulating in the FS monitor topic (left y-axis) spur Trigger invocations that start transfers to replicate automatically data between FSs. 
  } 
  \label{fig:data-aggregation}
  
\end{figure}

Here we tackle the challenge of reliably and efficiently synchronizing multiple parallel file systems (FS), for example by ensuring that the creation of new files in one FS spurs the replication of those files to the others.
We developed for this purpose the hierarchical EDA depicted in \autoref{fig:implemented-edas-arch}.
This solution starts with a FS monitor, \textit{FSMon}, developed in prior work~\cite{paul2019fsmonitor}, that collects events from a parallel FS.
One instance of this monitor per FS publishes events to a local Kafka topic, from which a local aggregator selects important and unique events for publication to \fabric{}. 
Then, an \fabric{} trigger filters these events, using the pattern in \autoref{lst:eventbridge}, for file creation events. For each such event, it makes a request to the Globus Transfer service~\cite{chard2016globus} to initiate a transfer from the source to the destination FS.
The invocation of triggers to process FS events from an exemplary time period is shown in~\autoref{fig:data-aggregation}.

\subsection{Online Task Scheduling}\label{sec:ots}

Here we tackle the challenge of obtaining timely information about tasks and resources in order to perform efficient online task scheduling.
To this end, we constructed an EDA that employs \fabric{} to communicate resource power and utilization information from remote resources to a FaaS scheduler~\cite{kamatar2024greenfaas}. 
Each managed resource has a Python-based monitor utilizing the Intel RAPL energy monitor~\cite{david2010rapl} and Python \textit{psutil} library~\cite{psutil} to collect resource usage information, which is then published to \fabric{}. The scheduler consumes this information to guide subsequent task placement and to train performance prediction models. 
\fabric{} provides an enabling communication layer for near real-time insight into the ongoing power usage of distributed resources. 

\subsection{Epidemic Modeling and Response}\label{sec:epi}

Researchers have adopted an EDA for a real-time epidemic monitoring and response platform~\cite{Eadline_2024, aero}. This system monitors various web-based data sources (e.g., public health data), and when data are updated, it ingests, cleans, and validates the data. Prediction models are regularly retrained and run, and data and model results are published for decision makers. 
The platform uses \fabric{} to manage the various events that trigger modeling. \fabric{} triggers are utilized to execute cloud-hosted workflows for data ingestion and model execution.  
Events include timer-based events to retrieve updates periodically from the various data sources; detection of anomalies; data updates; and new model results. 

\subsection{Dynamic Workflow Management}\label{sec:dwm}

Here we tackle the challenge of monitoring the activities of large parallel computations.
We work with Parsl~\cite{babuji2019parsl}, a parallel scripting library for Python that includes monitoring software to capture task execution and performance information from remote workers and record them in a centralized database. 
We extended Parsl with an EDA monitor using \fabric{} to capture global performance information. The \fabric{}-based monitoring tool publishes task and resource information, as well as task failure events. In future work, we will extend Parsl to use this information in various ways, for example, by retrying failed tasks, blacklisting under-performing nodes, or elastically rescheduling tasks when resource requirements exceed worker capabilities.

\begin{figure*}
    \centering
    \includegraphics[width=\textwidth,trim=1mm 1mm 1mm 1mm,clip]{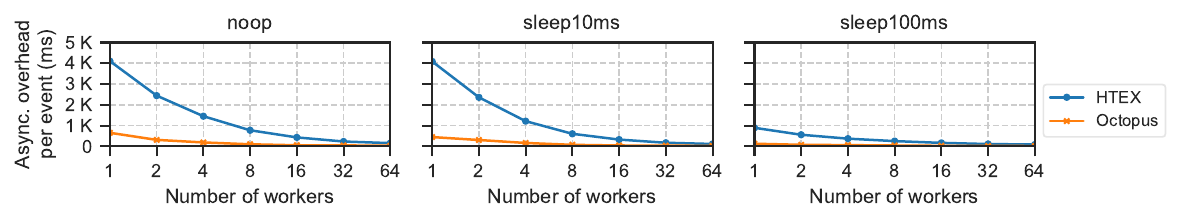}
    \label{fig:async-overhead}
    \vspace{-2em}
  \caption{Parsl workflow monitoring using the HTEX and \fabric{} systems to capture events up to 64 workers.}
  \vspace{-1.5em}
  \label{fig:exp-faster}
\end{figure*}

We ran scaling tests on FASTER~\cite{FASTER}, performing 128 tasks across eight nodes, varying the number of workers from 1 to 64 and task duration between 0, 10, and 100~ms. We calculate the overhead of each experiment by subtracting the task execution time from the total makespan, which includes the time to gather and log monitoring metrics, and then divide by the number of events generated in the experiment to determine the per-event cost. The results, in~\autoref{fig:exp-faster}, show that the average overhead per event decreases as the number of workers, and thus events, increases. This is due to the relatively static cost of writing events to a database. We see improved scalability with~\fabric{} due to its ability to batch events and publish them asynchronously.

\section{Discussion} \label{sec:discussion}
We discuss our experiences applying EDA, limitations of the approach, and costs of running \fabric{} as a cloud service.

\subsection{Scientific EDA Experiences}

To summarize our experiences applying an EDA to scientific applications, we reflect on the questions posed in~\autoref{sec:intro}.

\textit{How well do scientific applications map to EDA constructs?}

We observe that the growing reliance on data and computation resources that span locations and administrative domains creates challenges that are not well addressed by current approaches. However, this distribution aligns naturally with EDA, and particularly with its ability to manage and react asynchronously to distributed events. We find further that EDA provides the modularity required by applications composed of interconnected components or services, each responsible for different aspects of a workflow. EDA allows developers to define these individual components and then enable them to communicate and coordinate asynchronously through events, without tight coupling. In summary, the use cases considered here suggest that scientific applications with distributed workloads, real-time processing requirements, and component-based workloads map well to the EDA model.

\textit{What are the performance characteristics of different implementation approaches?} 
We evaluated various configurations of our prototype implementation to measure performance characteristics and found an approach that exceeds the requirements of our use cases and is also scalable for future requirements. The \fabric{} architecture provides minimal friction between clients and the event fabric while still providing a robust authentication and authorization model. Combined, these characteristics give \fabric{} the ability to manage millions of events each second, scale to accommodate future needs, and securely integrate with distributed research infrastructure.

\textit{How can we enable actions to be automatically launched as a result of events?}
We implemented a trigger-action programming model as it is both intuitive and flexible. \fabric{} Triggers leverage a well-known event filtering language and employ managed, scalable infrastructure to automatically process events, eliminating the need to manually operate consumers for real-time responsiveness.  

\textit{How feasible is it for EDA to integrate with existing DRI and scientific services to serve the research community?}
We found authentication and authorization to be the biggest integration challenge. It is important to leverage an existing federated identity solution that has been widely adopted across research computing centers. We build upon the Globus Auth ecosystem by implementing~\fabric{} as an OAuth resource server. Adoption of a common action API enables easy invocation of external actions while also simplifying the integration of new actions into the \fabric{} ecosystem.

\textit{How can we build a multi-tenant EDA that serves the needs and meets the scale of broad science use cases?}
We have evaluated and used~\fabric{} to implement several representative EDAs in many environments, including Cloud (Scheduling), Edge (Epidemic), and HPC (Data Automation and Workflow). Our studies support our contributions, and our evaluations include configurations that represent large deployments of scientific applications. For example, we test with rates exceeding 10,000 events/second and 100 event consumers running at full capacity (\autoref{table:evaluation}). These results are relevant for HPC clusters with thousands of computational nodes because Octopus is intended to be used through hierarchical aggregation (see Data Automation application), producing batched events sent to the cloud.

Our approach, building on scalable cloud platforms and research services, provides confidence that \fabric{} can scale to meet the needs of diverse scientific applications with varying data, throughput, and latency requirements.  The adoption of well-defined APIs ensures that \fabric{} can be extended to address future requirements.

\subsection{Limitations}
We also observed limitations of EDA that will require further investigation. 
\textbf{Complex event-driven applications} can be difficult to program, debug, and reason about, due to the chaining of asynchronous actions. Researchers have observed similar challenges in other domains such as smart homes, and proposed methods to improve the correctness of trigger-action programs~\cite{zhang23tap}.
\textbf{Failure detection} is complicated by the asynchronous nature of EDA. For example, capturing errors and determining the sequence of events that lead to a failure can be difficult to trace. Building fault-tolerant applications can also be challenging as actions may not be idempotent.
\textbf{Large-scale} systems, such as HPC systems, can overwhelm \fabric{} and the networks that connect it to the edge with their scale and rate of event generation.
Instead, hierarchical methods (like our approach with file system events) can be used to aggregate important events to the cloud and high-performance implementations of Kafka could be used to capture events at high rates. 
\textbf{Network partitions} between \fabric{}' cloud service and producers/consumers may render the system unusable. Caching can be applied to ensure that events are not lost, yet EDA-based applications, as with any cloud-hosted application, will be unusable during the partition. 

\subsection{Costs}

Our use of AWS services to host the \fabric{} platform incurs cloud provider costs. The MSK service requires a minimum of two nodes, with the smallest available node type costing \$0.0456 per hour for a minimum monthly cost of $\sim$\$70. Data egress and trigger processing can be more costly, with data egress from MSK to remote consumers at \$0.09 per GB and the Lambda service used for triggers at \$10 for 1 M requests (128~MB memory with 5~s duration). Thus, for example, a scheduling application (\autoref{table:use-cases}) processing \num{10000} events per hour for each of 10 resources would invoke 10,000$\times$10$\times$24 = 2.4 M lambdas per day, which if using a 5~s trigger and 4~KB events, costs \$24 daily. The incurred egress costs in this example would be negligible.  

Although these costs would be excessive in many settings, they can easily be reduced greatly. The \fabric{} service can rate limit invocations on a per-identity basis, restrict runtime configuration to minimize resource footprint, and promote proper event filtering by using EventBridge to reduce the number of trigger invocations.  The application can, as in \autoref{sec:dwm}, employ aggregators at resources to reduce trigger invocations by orders of magnitude. Any service could use these and other methods (e.g., rate-limiting consumers, compression, proxies) to manage costs.

\section{Related work} \label{sec:related-work}
This project operates at the intersection of HPC and distributed computing. Here we describe the current state-of-the-art and explain how \fabric{} builds upon existing efforts.

\textbf{Event-driven architectures}:
Message-oriented middleware was designed to address the challenges of directly and reliably communicating among increasingly distributed and heterogeneous resources by employing intermediary queues between producers and consumers~\cite{curry2004message}.
EDAs~\cite{michelson2006event} have been widely adopted in recent years~\cite{cabane2024impact} and often find success in facilitating large-scale~\cite{vijayarenu2020scaling} and real-time~\cite{leners2011detecting, lavanya2020real} applications due to their ability to process billions of events each minute. Further, the modularity of EDA can decompose monolithic applications into microservices to improve maintainability~\cite{laigner2020monolithic}.

The popularity of EDA has resulted in several managed cloud-native services, including Amazon Kinesis \cite{kinesisweb}, Google Cloud Pub/Sub \cite{pubsubweb}, and Azure Event Hubs \cite{eventhubsweb}, each deeply integrated with their respective ecosystems. \fabric{} builds on these services by using AWS MSK and extends it to create a multi-user EDA ecosystem that meets the needs of researchers and distributed research infrastructure.

\textbf{Event-driven science}:
The high data rates and inherent distribution of science necessitate event filtering and aggregation at the edge~\cite{alessio2008lhcb}. Researchers at the Large Hadron Collider (LHC) use an EDA to process over one million events per second~\cite{alessio2008lhcb}, with the first filtering layer performing a 1/500 data reduction. The remaining events are then passed to other services for analysis, monitoring, and storage. These rates demonstrate the need for hierarchical filtering. Hierarchical filtering has also proven useful in other contexts, such as federated learning communication~\cite{bano2022kafkafed} and provides an area of future work to determine how such practices should be applied in HPC settings.


The application of EDA within scientific contexts can often be found in data management and workflow tools~\cite{herath2010streamflow,pruyne2024steering}.
StreamFlow~\cite{herath2010streamflow} aims to integrate data streams into scientific workflows without altering the control flows. This model is designed to be deployed locally and does not serve highly distributed clients.
The integrated Rule-Oriented Data System (iRODS) \cite{rajasekar2010irods} is a prominent rule-based EDA for data management, sharing, publication, and long-term preservation of distributed data. While iRODS is not directly comparable to \fabric, the rules engine is insightful for trigger conditions.

Trigger-action programming, commonly used in home automation, for example via the If-This-Then-That service~\cite{ifttt}, provides simple interfaces for users to express conditional logic on events. Such concepts have also been applied to scientific data management~\cite{chard2017ripple}, allowing data events to spur the invocation of analysis scripts.
In contrast, \fabric{} offers flexibility in handling events by supporting both direct event consumption and the deployment of triggers.

\section{Summary} \label{sec:summary}
Modern science increasingly requires synchronized utilization of cutting-edge computing, networking, instruments, and experimental facilities, collectively referred to as distributed research infrastructure. These resources and their applications can generate many events, and as many science applications span locations, scientists need to consume events from many sources. To address these requirements, we proposed a novel, hybrid cloud-edge event fabric,~\fabric{}, with the scalability and flexibility needed to implement scientific EDAs. Results from both microbenchmarks and realistic application studies show that \fabric{} is able to meet the throughput, latency, and resilience needs of several real-world applications.  Based on our experience implementing scientific EDAs, we conclude that EDA constructs map well to scientific settings.

Our analysis of scientific EDAs and evaluation of \fabric{} also identified several areas for future work. The scale of HPC resources makes hierarchical event aggregation patterns a necessity and requires additional investigation into how such capabilities can be generally applied within HPC contexts. Research is also required into how complex event-driven applications can be reasoned about to minimize failures and aid in debugging. Further, the rapid identification of failures within an EDA is needed for their use as critical infrastructure.

\section*{Acknowledgment}

We thank the entire team of the Diaspora Project for their helpful comments and feedback. This material is based upon work supported by the U.S. Department of Energy (DOE), Office of Science, Office of Advanced Scientific Computing Research, under Contract DE-AC02-06CH11357.

\small
\bibliographystyle{IEEEtran}
\bibliography{bibs/references}

\begin{thebibliography}{10}
\providecommand{\url}[1]{#1}
\csname url@samestyle\endcsname
\providecommand{\newblock}{\relax}
\providecommand{\bibinfo}[2]{#2}
\providecommand{\BIBentrySTDinterwordspacing}{\spaceskip=0pt\relax}
\providecommand{\BIBentryALTinterwordstretchfactor}{4}
\providecommand{\BIBentryALTinterwordspacing}{\spaceskip=\fontdimen2\font plus
\BIBentryALTinterwordstretchfactor\fontdimen3\font minus
  \fontdimen4\font\relax}
\providecommand{\BIBforeignlanguage}[2]{{%
\expandafter\ifx\csname l@#1\endcsname\relax
\typeout{** WARNING: IEEEtran.bst: No hyphenation pattern has been}%
\typeout{** loaded for the language `#1'. Using the pattern for}%
\typeout{** the default language instead.}%
\else
\language=\csname l@#1\endcsname
\fi
#2}}
\providecommand{\BIBdecl}{\relax}
\BIBdecl

\bibitem{IRI}
{ASCR Integrated Research Infrastructure Task Force}, ``Toward a seamless
  integration of computing, experimental, and observational science facilities:
  A blueprint to accelerate discovery,'' US DOE Office of Science, Advanced
  Scientific Computing Research, Tech. Rep., 2021,
  \url{http://doi.org/10.2172/1863562}.

\bibitem{babu2023deep}
A.~V. Babu, T.~Zhou, S.~Kandel, T.~Bicer, Z.~Liu, W.~Judge, D.~J. Ching,
  Y.~Jiang, S.~Veseli, S.~Henke, R.~Chard, Y.~Yao, E.~Sirazitdinova, G.~Gupta,
  M.~V. Holt, I.~T. Foster, A.~Miceli, and M.~J. Cherukara, ``Deep learning at
  the edge enables real-time streaming ptychographic imaging,'' \emph{Nature
  Communications}, vol.~14, no.~1, p. 7059, 2023.

\bibitem{pithan2023closing}
L.~Pithan, V.~Starostin, D.~Mare{\v{c}}ek, L.~Petersdorf, C.~V{\"o}lter,
  V.~Munteanu, M.~Jankowski, O.~Konovalov, A.~Gerlach, A.~Hinderhofer,
  B.~Murphy, S.~Kowarik, and F.~Schreiber, ``Closing the loop: Autonomous
  experiments enabled by machine-learning-based online data analysis in
  synchrotron beamline environments,'' \emph{Journal of Synchrotron Radiation},
  vol.~30, no.~6, 2023.

\bibitem{sparkes2010towards}
A.~Sparkes, W.~Aubrey, E.~Byrne, A.~Clare, M.~N. Khan, M.~Liakata, M.~Markham,
  J.~Rowland, L.~N. Soldatova, K.~E. Whelan, M.~Young, and R.~D. King,
  ``Towards robot scientists for autonomous scientific discovery,''
  \emph{Automated Experimentation}, vol.~2, pp. 1--11, 2010.

\bibitem{hase2019next}
F.~H{\"a}se, L.~M. Roch, and A.~Aspuru-Guzik, ``Next-generation experimentation
  with self-driving laboratories,'' \emph{Trends in Chemistry}, vol.~1, no.~3,
  pp. 282--291, 2019.

\bibitem{stach2021autonomous}
E.~Stach, B.~DeCost, A.~G. Kusne, J.~Hattrick-Simpers, K.~A. Brown, K.~G.
  Reyes, J.~Schrier, S.~Billinge, T.~Buonassisi, I.~Foster, , C.~P. Gomes,
  J.~M. Gregoire, A.~Mehta, J.~Montoya, E.~Olivetti, C.~Park, E.~Rotenberg,
  S.~K. Saikin, S.~Smullin, V.~Stanev, and B.~Maruyama, ``Autonomous
  experimentation systems for materials development: A community perspective,''
  \emph{Matter}, vol.~4, no.~9, pp. 2702--2726, 2021.

\bibitem{abolhasani2023rise}
M.~Abolhasani and E.~Kumacheva, ``The rise of self-driving labs in chemical and
  materials sciences,'' \emph{Nature Synthesis}, vol.~2, no.~6, pp. 483--492,
  2023.

\bibitem{collier2023developing}
N.~Collier, J.~M. Wozniak, A.~Stevens, Y.~Babuji, M.~Binois, A.~Fadikar,
  A.~W{\"u}rth, K.~Chard, and J.~Ozik, ``Developing distributed
  high-performance computing capabilities of an open science platform for
  robust epidemic analysis,'' in \emph{IEEE International Parallel and
  Distributed Processing Symposium Workshops}.\hskip 1em plus 0.5em minus
  0.4em\relax IEEE, 2023, pp. 868--877.

\bibitem{vescovi2022linking}
R.~Vescovi, R.~Chard, N.~D. Saint, B.~Blaiszik, J.~Pruyne, T.~Bicer, A.~Lavens,
  Z.~Liu, M.~E. Papka, S.~Narayanan, N.~Schwarz, K.~Chard, and I.~T. Foster,
  ``Linking scientific instruments and computation: Patterns, technologies, and
  experiences,'' \emph{Patterns}, vol.~3, no.~10, p. 100606, 2022.

\bibitem{chard23automation}
\BIBentryALTinterwordspacing
R.~Chard, J.~Pruyne, K.~McKee, J.~Bryan, B.~Raumann, R.~Ananthakrishnan,
  K.~Chard, and I.~T. Foster, ``Globus automation services: Research process
  automation across the space–time continuum,'' \emph{Future Generation
  Computer Systems}, vol. 142, pp. 393--409, 2023. [Online]. Available:
  \url{https://www.sciencedirect.com/science/article/pii/S0167739X23000183}
\BIBentrySTDinterwordspacing

\bibitem{michelson2006event}
B.~M. Michelson, ``Event-driven architecture overview,'' Patricia Seybold
  Group, {Tech Report}, 2006,
  \url{https://complexevents.com/wp-content/uploads/2006/07/OMG-EDA-bda2-2-06cc.pdf}.

\bibitem{hinze2009event}
A.~Hinze, K.~Sachs, and A.~Buchmann, ``Event-based applications and enabling
  technologies,'' in \emph{3rd ACM International Conference on Distributed
  Event-Based Systems}, 2009, pp. 1--15.

\bibitem{dayarathna2018recent}
M.~Dayarathna and S.~Perera, ``Recent advancements in event processing,''
  \emph{ACM Computing Surveys}, vol.~51, no.~2, pp. 1--36, 2018.

\bibitem{huang15tap}
\BIBentryALTinterwordspacing
J.~Huang and M.~Cakmak, ``Supporting mental model accuracy in trigger-action
  programming,'' in \emph{ACM International Joint Conference on Pervasive and
  Ubiquitous Computing}, ser. UbiComp '15.\hskip 1em plus 0.5em minus
  0.4em\relax New York, NY, USA: Association for Computing Machinery, 2015, p.
  215–225. [Online]. Available: \url{https://doi.org/10.1145/2750858.2805830}
\BIBentrySTDinterwordspacing

\bibitem{ur16trigger}
\BIBentryALTinterwordspacing
B.~Ur, M.~Pak Yong~Ho, S.~Brawner, J.~Lee, S.~Mennicken, N.~Picard, D.~Schulze,
  and M.~L. Littman, ``Trigger-action programming in the wild: An analysis of
  200,000 ifttt recipes,'' in \emph{Conference on Human Factors in Computing
  Systems}, ser. CHI '16.\hskip 1em plus 0.5em minus 0.4em\relax New York, NY,
  USA: Association for Computing Machinery, 2016, p. 3227–3231. [Online].
  Available: \url{https://doi.org/10.1145/2858036.2858556}
\BIBentrySTDinterwordspacing

\bibitem{seifrid22sdl}
M.~Seifrid, R.~Pollice, A.~Aguilar-Granda, Z.~Morgan~Chan, K.~Hotta, C.~T. Ser,
  J.~Vestfrid, T.~C. Wu, and A.~Aspuru-Guzik, ``Autonomous chemical
  experiments: Challenges and perspectives on establishing a self-driving
  lab,'' \emph{Accounts of Chemical Research}, vol.~55, no.~17, pp. 2454--2466,
  2022.

\bibitem{gray2005scientific}
J.~Gray, D.~T. Liu, M.~Nieto-Santisteban, A.~Szalay, D.~J. DeWitt, and
  G.~Heber, ``Scientific data management in the coming decade,'' \emph{Acm
  Sigmod Record}, vol.~34, no.~4, pp. 34--41, 2005.

\bibitem{rajasekar2010irods}
A.~Rajasekar, R.~Moore, C.-y. Hou, C.~A. Lee, R.~Marciano, A.~de~Torcy, M.~Wan,
  W.~Schroeder, S.-Y. Chen, L.~Gilbert, P.~Tooby, and B.~Zhu, ``{iRODS Primer}:
  Integrated rule-oriented data system,'' \emph{Synthesis Lectures on
  Information Concepts, Retrieval, and Services}, vol.~2, no.~1, pp. 1--143,
  2010.

\bibitem{chard2020funcx}
R.~Chard, Y.~Babuji, Z.~Li, T.~Skluzacek, A.~Woodard, B.~Blaiszik, I.~Foster,
  and K.~Chard, ``Func{X}: A federated function serving fabric for science,''
  in \emph{29th International Symposium on High-performance Parallel and
  Distributed Computing}, 2020, pp. 65--76.

\bibitem{o2017digital}
J.~O'Shea, ``Digital disease detection: A systematic review of event-based
  internet biosurveillance systems,'' \emph{International Journal of Medical
  Informatics}, vol. 101, pp. 15--22, 2017.

\bibitem{strohbach2015towards}
M.~Strohbach, H.~Ziekow, V.~Gazis, and N.~Akiva, ``Towards a big data analytics
  framework for {IoT} and smart city applications,'' in \emph{Modeling and
  processing for next-generation big-data technologies: With applications and
  case studies}.\hskip 1em plus 0.5em minus 0.4em\relax Springer, 2015, pp.
  257--282.

\bibitem{catlett20aot}
C.~Catlett, P.~Beckman, N.~Ferrier, H.~Nusbaum, M.~E. Papka, M.~G. Berman, and
  R.~Sankaran, ``Measuring cities with software-defined sensors,''
  \emph{Journal of Social Computing}, vol.~1, no.~1, pp. 14--27, 2020.

\bibitem{altintas23wifire}
\BIBentryALTinterwordspacing
I.~Altintas, J.~Block, D.~L. Crawl, and R.~A.~d. Callafon, \emph{Using Dynamic
  Data-Driven Cyberinfrastructure for Next-Generation Wildland Fire
  Intelligence}.\hskip 1em plus 0.5em minus 0.4em\relax Cham: Springer
  International Publishing, 2023, pp. 451--474. [Online]. Available:
  \url{https://doi.org/10.1007/978-3-031-27986-7_17}
\BIBentrySTDinterwordspacing

\bibitem{li22faulttolerant}
\BIBentryALTinterwordspacing
C.~Li, J.~Liu, M.~Wang, and Y.~Luo, ``Fault-tolerant scheduling and data
  placement for scientific workflow processing in geo-distributed clouds,''
  \emph{Journal of Systems and Software}, vol. 187, p. 111227, 2022. [Online].
  Available:
  \url{https://www.sciencedirect.com/science/article/pii/S0164121222000073}
\BIBentrySTDinterwordspacing

\bibitem{kreps2011kafka}
J.~Kreps, N.~Narkhede, and J.~Rao, ``Kafka: A distributed messaging system for
  log processing,'' in \emph{Proceedings of the NetDB}, vol.~11, 2011, pp.
  1--7.

\bibitem{linkedin2019how}
J.~Lee, ``How {LinkedIn} customizes {Apache} {Kafka} for 7 trillion messages
  per day,''
  \url{https://www.linkedin.com/blog/engineering/open-source/apache-kafka-trillion-messages}.
  Accessed September 2024.

\bibitem{tuecke2016globus}
S.~Tuecke, R.~Ananthakrishnan, K.~Chard, M.~Lidman, B.~McCollam, S.~Rosen, and
  I.~Foster, ``Globus {A}uth: A research identity and access management
  platform,'' in \emph{IEEE 12th International Conference on e-Science}.\hskip
  1em plus 0.5em minus 0.4em\relax IEEE, 2016, pp. 203--212.

\bibitem{hunt2010zookeeper}
P.~Hunt, M.~Konar, F.~P. Junqueira, and B.~Reed, ``{ZooKeeper}: Wait-free
  coordination for internet-scale systems,'' in \emph{USENIX Annual Technical
  Conference}, 2010.

\bibitem{zookeeper}
{The Apache Software Foundation}, ``{Apache ZooKeeper},''
  \url{https://zookeeper.apache.org}. Accessed September 2024.

\bibitem{eventbridgepatterns}
{Amazon Web Services}, ``{Content filtering in Amazon EventBridge event
  patterns},''
  \url{https://docs.aws.amazon.com/eventbridge/latest/userguide/eb-pipes-event-filtering.html}.
  Accessed September 2024.

\bibitem{paul2019fsmonitor}
A.~K. Paul, R.~Chard, K.~Chard, S.~Tuecke, A.~R. Butt, and I.~Foster,
  ``Fsmonitor: Scalable file system monitoring for arbitrary storage systems,''
  in \emph{2019 IEEE International Conference on Cluster Computing
  (CLUSTER)}.\hskip 1em plus 0.5em minus 0.4em\relax IEEE, 2019, pp. 1--11.

\bibitem{chard2016globus}
K.~Chard, S.~Tuecke, and I.~Foster, ``Globus: Recent enhancements and future
  plans,'' in \emph{XSEDE16 Conference on Diversity, Big Data, and Science at
  Scale}.\hskip 1em plus 0.5em minus 0.4em\relax ACM, 2016, p.~27.

\bibitem{kamatar2024greenfaas}
\BIBentryALTinterwordspacing
A.~Kamatar, V.~Hayot-Sasson, Y.~Babuji, A.~Bauer, G.~Rattihalli, N.~Hogade,
  D.~Milojicic, K.~Chard, and I.~Foster, ``Greenfaas: Maximizing energy
  efficiency of hpc workloads with faas,'' 2024. [Online]. Available:
  \url{https://arxiv.org/abs/2406.17710}
\BIBentrySTDinterwordspacing

\bibitem{david2010rapl}
H.~David, E.~Gorbatov, U.~R. Hanebutte, R.~Khanna, and C.~Le, ``{RAPL}: Memory
  power estimation and capping,'' in \emph{16th ACM/IEEE International
  Symposium on Low Power Electronics and Design}, 2010, pp. 189--194.

\bibitem{psutil}
G.~Rodola, ``psutil cross-platform library for process and system monitoring in
  {P}ython,'' \url{https://pypi.org/project/psutil/}. Accessed September 2024.

\bibitem{Eadline_2024}
\BIBentryALTinterwordspacing
D.~Eadline, ``Preparing for the next pandemic: Developing an open science
  platform for better decision-making in public health,'' Aug 2024. [Online].
  Available: \url{https://bit.ly/47COAd5}
\BIBentrySTDinterwordspacing

\bibitem{aero}
V.~Hayot-Sasson, A.~Stevens, N.~Collier, S.~Sridhar, K.~Conroy, J.~G. Pauloski,
  Y.~Babuji, M.~Gonthier, D.~D. Sanchez-Gallegos, I.~Foster, K.~Chard, and
  J.~Ozik, ``{AERO: An Autonomous Platform for Continuous Research},''
  \emph{ArXiv Preprint}, 2024.

\bibitem{babuji2019parsl}
Y.~Babuji, A.~Woodard, Z.~Li, D.~S. Katz, B.~Clifford, R.~Kumar, L.~Lacinski,
  R.~Chard, J.~M. Wozniak, I.~Foster \emph{et~al.}, ``Parsl: Pervasive parallel
  programming in {P}ython,'' in \emph{28th International Symposium on
  High-Performance Parallel and Distributed Computing}, 2019, pp. 25--36.

\bibitem{FASTER}
{TAMU HPRC, Texas A\&M University}, ``Texas {A\&M} high performance research
  computing,'' \url{https://hprc.tamu.edu/faster}. Accessed September 2024.

\bibitem{zhang23tap}
\BIBentryALTinterwordspacing
L.~Zhang, C.~Zhou, M.~L. Littman, B.~Ur, and S.~Lu, ``Helping users debug
  trigger-action programs,'' \emph{Proceedings of the ACM on Interactive,
  Mobile, Wearable and Ubiquitous Technologies}, vol.~6, no.~4, jan 2023.
  [Online]. Available: \url{https://doi.org/10.1145/3569506}
\BIBentrySTDinterwordspacing

\bibitem{curry2004message}
E.~Curry, ``Message-oriented middleware,'' \emph{Middleware for
  communications}, pp. 1--28, 2004.

\bibitem{cabane2024impact}
H.~Cabane and K.~Farias, ``On the impact of event-driven architecture on
  performance: An exploratory study,'' \emph{Future Generation Computer
  Systems}, vol. 153, pp. 52--69, 2024.

\bibitem{vijayarenu2020scaling}
L.~VijayaRenu, Z.~Wang, and J.~Rottinghuis, ``Scaling event aggregation at
  {T}witter to handle billions of events per minute,'' in \emph{2020 IEEE
  Infrastructure Conference}.\hskip 1em plus 0.5em minus 0.4em\relax IEEE,
  2020, pp. 1--4.

\bibitem{leners2011detecting}
J.~B. Leners, H.~Wu, W.-L. Hung, M.~K. Aguilera, and M.~Walfish, ``Detecting
  failures in distributed systems with the {F}alcon spy network,'' in
  \emph{23rd ACM Symposium on Operating Systems Principles}, 2011, pp.
  279--294.

\bibitem{lavanya2020real}
K.~Lavanya, S.~Venkatanarayanan, and A.~A. Bhoraskar, ``Real-time weather
  analytics: An end-to-end big data analytics service over {Apache Spark with
  Kafka} and long short-term memory networks,'' \emph{International Journal of
  Web Services Research (IJWSR)}, vol.~17, no.~4, pp. 15--31, 2020.

\bibitem{laigner2020monolithic}
R.~Laigner, M.~Kalinowski, P.~Diniz, L.~Barros, C.~Cassino, M.~Lemos,
  D.~Arruda, S.~Lifschitz, and Y.~Zhou, ``From a monolithic big data system to
  a microservices event-driven architecture,'' in \emph{2020 46th Euromicro
  conference on software engineering and advanced applications (SEAA)}.\hskip
  1em plus 0.5em minus 0.4em\relax IEEE, 2020, pp. 213--220.

\bibitem{kinesisweb}
{Amazon Web Services}, ``Data {Stream} {Processing} - {Amazon} {Kinesis} --
  {AWS},'' \url{https://aws.amazon.com/kinesis}. Accessed September 2024.

\bibitem{pubsubweb}
{Google Cloud}, ``Pub/{Sub} for {Application} \& {Data} {Integration},''
  \url{https://cloud.google.com/pubsub}. Accessed September 2024.

\bibitem{eventhubsweb}
{Microsoft}, ``Event {Hubs}—{Real}-{Time} {Data} {Ingestion} {\textbar}
  {Microsoft} {Azure},''
  \url{https://azure.microsoft.com/en-us/products/event-hubs}. Accessed
  September 2024.

\bibitem{alessio2008lhcb}
F.~Alessio, C.~Barandela, L.~Brarda, M.~Frank, B.~Franek, D.~Galli, C.~Gaspar,
  E.~v~Herwijnen, R.~Jacobsson, B.~Jost \emph{et~al.}, ``Lhcb online event
  processing and filtering,'' in \emph{Journal of Physics: Conference Series},
  vol. 119.\hskip 1em plus 0.5em minus 0.4em\relax IOP Publishing, 2008, p.
  022003.

\bibitem{bano2022kafkafed}
S.~Bano, N.~Tonellotto, P.~Cassar{\`a}, and A.~Gotta, ``Kafkafed: Two-tier
  federated learning communication architecture for internet of vehicles,'' in
  \emph{2022 IEEE International Conference on Pervasive Computing and
  Communications Workshops and other Affiliated Events (PerCom
  Workshops)}.\hskip 1em plus 0.5em minus 0.4em\relax IEEE, 2022, pp. 515--520.

\bibitem{herath2010streamflow}
C.~Herath and B.~Plale, ``Streamflow programming model for data streaming in
  scientific workflows,'' in \emph{2010 10th IEEE/ACM International Conference
  on Cluster, Cloud and Grid Computing}.\hskip 1em plus 0.5em minus 0.4em\relax
  IEEE, 2010, pp. 302--311.

\bibitem{pruyne2024steering}
J.~Pruyne, V.~Hayot-Sasson, W.~Zheng, R.~Chard, J.~M. Wozniak, T.~Bicer,
  K.~Chard, and I.~T. Foster, ``Steering a fleet: Adaptation for large-scale,
  workflow-based experiments,'' 2024.

\bibitem{ifttt}
L.~Tibbets, ``{If This Then That},'' \url{https://www.ifttt.com}. Accessed
  September 2024.

\bibitem{chard2017ripple}
R.~Chard, K.~Chard, J.~Alt, D.~Y. Parkinson, S.~Tuecke, and I.~Foster,
  ``Ripple: Home automation for research data management,'' in \emph{37th
  International Conference on Distributed Computing Systems Workshops}.\hskip
  1em plus 0.5em minus 0.4em\relax IEEE, 2017, pp. 389--394.

\end{thebibliography}

\end{document}